# CONTINUUM POWER SPECTRUM COMPONENTS IN X–RAY SOURCES: DETAILED MODELLING AND SEARCH FOR COHERENT PERIODICITIES

L. Stella[*], E. Arlandi, G. Tagliaferri,

Osservatorio Astronomico di Brera, Via E. Bianchi 46, I-22055 Merate (Como), Italy

and

G.L. Israel[*]

International School for Advanced Studies (SISSA–ISAS), Via Beirut 2-4, I-34013 Trieste, Italy

**ABSTRACT**

This paper summarises two recently developed techniques in power spectral analysis and their application to a sample of X-ray light curves of accreting collapsed objects in active galactic nuclei and X-ray binaries. The first technique is designed to carry out detailed model fitting of continuum power spectrum components arising from noise variability by using maximum likelihood methods. The technique is applied to the light curves of a number of highly variable AGNs observed with EXOSAT. Substantially steeper logarithmic power spectrum slopes are obtained than previously estimated with standard methods. The second technique was devised in order to reveal coherent periodicities in the presence of "coloured" (i.e. non–white) noise variability components from the source. To this aim the power spectra are searched for significant narrow peaks superposed on the "coloured" continuum components. We present the results of a search for an orbital modulation in the light curves of a sample of 25 low mass X-ray binaries (LMXRBs), for which the orbital period is either unknown or detected only at optical wavelengths. This led to the discovery of a significant X-ray orbital modulation at the few percent level in the burster MXB 1636−539.

## 1 Introduction

Astronomical time series of increased statistical quality, time resolution and duration have become available in recent years for a variety of objects and over different bands of the electromagnetic spectrum. Power spectrum analysis is probably the single most important technique that is applied to these time series, in order to (a) detect periodicities and quasi-periodicities, by the presence of significant power spectrum peaks; (b) characterise the noise variability through the study of continuuum power spectrum components.

Applications to high energy astronomical time series have been especially numerous and successfull over the last decade, as a consequence of the pronounced variability (often both periodic and non-periodic in character) detected in many sources and the availability of long uninterrupted observations (up to several days) of high signal to noise ratio. Accreting collapsed stars such as those in X-ray binaries (neutron stars and stellar mass black holes) and active galactic nuclei (AGNs) ($10^6 - 10^9$ $M_\odot$ black holes) often display different energy spectrum and/or intensity states which are sometimes accompanied by different time variability properties. The continuum power spectral components arising from noise variability usually increase towards lower frequencies (*red noise*), often in a power law-like fashion. Their study has proven to be a useful tool for morphological classifications and, sometimes, has provided constraints on physical models (e.g. Stella 1988; Lewin, van Paradijs and van der Klis 1988; Hasinger and van der Klis 1989). To derive quantitative information it has become customary to carry out detailed fitting of power spectra using analytical models; in this sense power spectra are treated more and more in a fashion similar to energy spectra.

The periodic modulations that have been revealed in a number of high energy sources often arise from the rotation of a compact magnetic star, or the orbital motion of a binary system. The detection and accurate measurement of these periods provides a tool of paramount importance. For instance, in the early 1970s the measurements of the orbital period and the secular changes of the spin period in binary X–ray pulsars proved that the X–ray emission in these systems is powered by accretion and allowed to obtain the first measurements of neutron star masses. A variety of other periodic or quasi-periodic phenomena in X–ray sources have been discovered over a variety of timescales (from tens of milliseconds to years). Their interpretation ranges from, e.g., precession, to radial oscillations, accretion disk - magnetosphere interactions, motions or occultations in an accretion disk, activity of the companion star etc.

Astronomical observations at X–ray and $\gamma$–ray energies rely upon photon counting instruments; therefore, measurement errors are often dominated by statistical uncertainties originating from the Poisson distribution of the counts. This translates into a white noise power spectrum component of known amplitude, which, af-degrees of freedom ($\chi_2^2$). Any intrinsic variability of the source, either resulting from periodic signal(s) or from noise(s), must possess significant power above the counting statistics white noise component in order to be detected.

## 2 Two problems in the study of "coloured" power spectra

A typical approach to the study of power spectra characterised by continuum components consists in dividing the time series from a long observation in a number of consecutive intervals (usually $M > 20$) and obtaining an average sample spectrum from the sample spectra calculated from individual intervals. This procedure presents the following advantages: (i) it decreases the variance of the estimates of the power for individual Fourier frequencies; (ii) it ensures that the distribution of these estimates can be approximated by a Gaussian distribution, the variance of which can be obtained over the ensemble of the intervals; (iii) it allows to use least squares techniques for model fitting (which require the estimates to follow a Gaussian distribution). However, when compared to a single sample spectrum calculated over the entire duration of the observation, this approach reduces the Fourier resolution by a factor of M. This limits the low frequency range of the sample spectrum and reduces the sensitivity to narrow features, such as the peaks arising from a periodic signal. Therefore, the main advantages of long and uninterrupted observations are lost.

The alternative approach of fitting the single sample spectrum calculated from the entire observation presents with problems related to the non-Gaussian distribution of the spectral estimates, which makes least squares model fitting unreliable. Yet a number of authors have carried out model fitting of single sample spectra that adopt standard least-square techniques (as evidence by the lack of any mention of alternative techniques). Based on the fact that for a very wide range of stationary processes the spectral estimates follow a rescaled $\chi^2$ distribution, we have developed a suitable maximum likelihood method to carry out detailed model fitting and parameter estimation of sample spectra. In Section 4 the new technique is briefly outlined and its results are compared with those from least square fitting of power spectra, in a few interesting cases (for a full description see Arlandi et al. 1995). Section 5 presents the application of the new technique to the long and uninterrupted X-ray light curves of bright AGNs observed with EXOSAT.

Traditionally, the detection of peak(s) in the sample spectrum has been carried out either (i) by eye, in all those cases in which the peak amplitude is so large that it is self-evident or (ii) by ruling out (at a given confidence level) that a peak originates from an underlying white noise. The latter technique implicitely assumes that the power spectra do not possess any conspicuous "coloured" component above the white noise. In many istances this hypothesis is not verified at least over a range of frequencies. Indeed the very presence of "coloured" continuum power spectrum components resulting from the noise variability of the source makes the detection of significant power spectrum peaks a difficult statistical problem.

In general, establishing whether or not a sample spectrum peak originates from a periodic modulation requires instead an evaluation of the peak significance with respect to the *local* continuum of the sample, which in turn, can be dominated by the aperiodic variability of the source. Techniques along these lines have been developed, which rely upon smoothing or incoherent summation of sample spectra, to decrease the variance of the power estimates and/or allow the use of relatively simple statistics. In this way, however, the frequency resolution and, correspondingly, the sensitivity of the searches is reduced (e.g. Jenkins and Watts 1968; van der Klis 1989). Moreover, standard spectral smoothing does not allow to reproduce power law-like spectral shapes with acceptable accuracy.

In section 6 we outline a new technique that allows to detect sample spectrum narrow peaks, in the presence of "coloured" power spectrum components, while preserving the highest Fourier resolution (for a full description see Israel and Stella 1995). In Section 7 we describe the application of the new technique to a search for an orbital modulation in the light curves of a sample of 25 X-ray binaries observed with EXOSAT.

## 3 The distribution of the power estimates

When "coloured" continuum power spectrum components resulting from source variability are present, the statistical distribution of the corresponding power estimates cannot be derived from first principles. In the presence of extensive and repeated observations, the statistical properties of these power spectrum components could be obtained directly from the data. In practice this is difficult to do, because of the limited duration of the observation and the characteristics rednoise spectra which are commonly found: a single sample spectrum is often calculated over the entire observation duration, $T$, in order to explore the lowest possible frequencies, while maintaining the highest Fourier resolution ($\Delta\nu_F = 1/T$). In this case only one power estimate is obtained for each Fourier frequency and the statistical distribution of the noise component(s) from the source remains unexplored.

Alternatively the observation can be divided in a series of $M$ consecutive intervals and the distribution of the power estimates investigated over the ensemble of the sample spectra from individual intervals of duration $T/M$. To illustrate this, we divided an 8 ms resolved ∼5.5 hr long light curve of the black hole candidate X–ray binary Cyg X-1 into $M = 1244$ intervals of 640 s and calculated the spectrum from each interval.

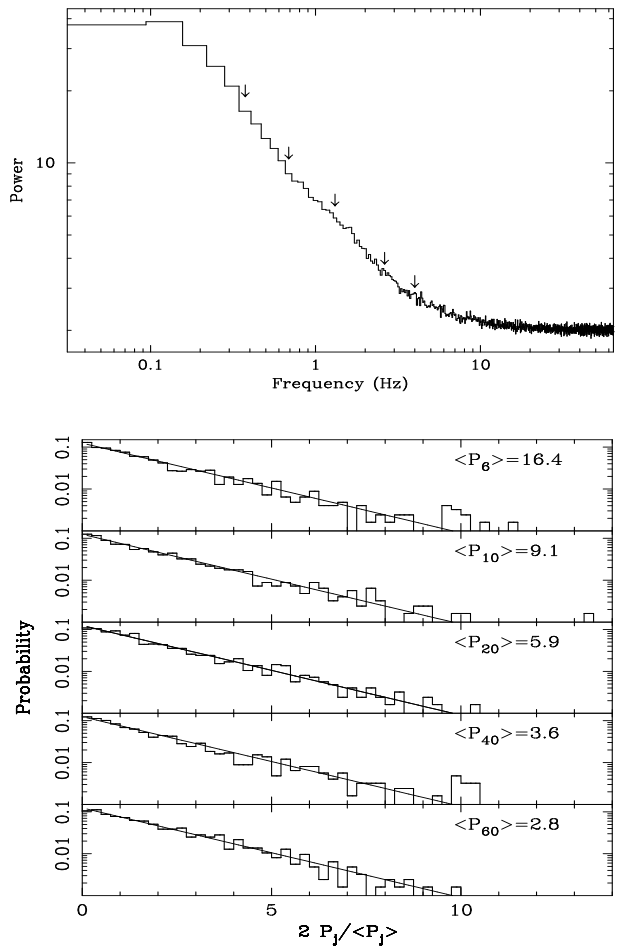

Figure 1: (a) Average sample spectrum from a 1-20 keV EXOSAT observation of the black hole candidate X-ray binary Cyg X-1. (b) Distribution of the normalised spectral estimates for selected Fourier frequencies ($j = 6, 10, 20, 40, 60$, see arrows in Fig. 1a). The solid lines represent a $\chi_2^2$-distribution.

Fig. 1 shows the distribution over the $M$ sample spectra of the power estimates for selected frequencies. Each distribution is normalised by $2/<P_j>$, where $<P_j>$ is the estimate of the average power at the $j-th$ Fourier frequency. It is apparent that in all cases the distribution is close to a $\chi_2^2$-distribution (also plotted for comparison). By extrapolating this result, we assume that, in general, the "coloured" noise components in the sample spectra of cosmic sources also follow a rescaled $\chi_2^2$-distribution. There is at least a very important class of random processes for which the power spectral estimates possess properties compatible with those discussed above. These are linear processes, $Y(t)$, in which a white noise, $Z(t)$ is passed through a linear filter $h(\tau)$, i.e.

$$Y(t) - \mu = \int_0^\infty h(\tau) Z(t-\tau)\, d\tau \qquad (1)$$

0. The power spectrum, $\Gamma_y$ of a linear process is given by:

$$\Gamma_y(\nu) = |H(\nu)|^2 \Gamma_z(\nu) \qquad (2)$$

where $\Gamma_z$ is the power spectrum of $Z(t)$ and $H(\nu)$ is the frequency response function of the linear filter $h(t)$. The power spectrum is the average over the realisations of the sample spectrum ($\Gamma_y(\nu) = E[Y(\nu)]$, $\Gamma_z(\nu) = E[Z(\nu)]$). Given a white noise source and a suitable linear filter it is then possible to generate a random process with arbitrary spectrum.

We assume that the "measurement" white noise component (Poisson noise in the case of counting detectors) can be interpreted as the input process $Z(t)$, such that Eq. (1) still holds. While clearly non-physical, this assumption involves no (statistical) approximation and allows to considerably simplify our treatment. In particular, it follows that for a given frequency $\nu$, the sample spectrum, $Y(\nu)$, of the linear process follows the same $\chi_2^2$-like distribution of the sample spectrum of the white noise, multiplied by $|H(\nu)|^2$. Therefore, the probability distribution of $Y(\nu_j)$ is (see also Arlandi et al. 1995; Israel and Stella 1995)

$$p_j(y, |H(\nu_j)|^2) = \frac{e^{-y/2|H(\nu_j)|^2}}{2|H(\nu_j)|^2} \qquad (3)$$

To study the continuum spectrum components in a computationally efficient way, $L$ contiguous power estimates at the Fourier frequencies are sometimes averaged towards high frequencies in order to attain a nearly constant relative resolution. This approach requires $L$ to increase for increasing frequencies and presents with the advantage of reducing the variance of the spectral estimates accordingly. Moreover, the sample spectrum can be the average of the spectra from $M$ different observations. In general, therefore, each spectral estimate can be the result of averaging $K = LM$ individual estimates and its statistical distribution is, therefore, related to a $\chi^2$-distribution with $2K$ degrees of freedom. This case is described in detail by Arlandi et al. (1995).

## 4 Maximum likelihood fitting of power spectra

Let $(\nu_1, \ldots, \nu_n)$ and $(y_1, \ldots, y_n)$ be the set of frequencies and the corresponding power estimates of the sample spectrum, and $f(\nu_j; \vec{a})$ the model function depending on a set of $p$ free parameters $\vec{a} = a_1, \ldots, a_p$. The model function $f(\nu_j; \vec{a})$ must represent, for each frequency $\nu_j$, the average over the ensemble of the realisations drawn from the distribution in Eq. 3,

$$\begin{aligned} f(\nu_j; \vec{a}) &= E[Y(\nu_j)] \\ &= \int_0^\infty y p_j(y, |H(\nu_j)|^2)\, dy = 2|H(\nu_j)|^2 \end{aligned} \qquad (4)$$

given by the likelihood function

$$P(y_1,\ldots,y_n,f(\nu_1;\vec{a}),\ldots,f(\nu_n;\vec{a}))$$
$$= \prod_{j=1}^{n} p_j(y_j,f(\nu_j;\vec{a})) = \prod_{j=1}^{n} \frac{e^{-\frac{y_j}{f(\nu_j;\vec{a})}}}{f(\nu_j;\vec{a})} \quad (5)$$

(above, we have considered the $y_1,\ldots,y_n$ as independently distributed, a good approximation for large $n$, see e.g. Jenkins and Watts 1968). The maximum likelihood estimate of the model function parameters $\vec{a}$ is obtained by maximizing the above probability, or, equivalently, by minimizing

$$-2\log P = 2\sum_{j=1}^{n}(\log f(\nu_j;\vec{a}) + \frac{y_j}{f(\nu_j;\vec{a})}) \quad (6)$$

The maximum likelihood model fitting and parameter estimation described above must be complemented with a method to evaluate the goodness of fit (see also Anderson et al. 1990). Unlike the Gaussian case, the minimum value of $-2\log P$ does not provide direct information on the goodness of fit (note that the first term in the sum does not contain the $y_j$) and can even be negative. However, if $f(\nu;\vec{a})$ provides a good fit then the ratio $\frac{2Y(\nu_j)}{f(\nu_j;\vec{a})}$ for each $\nu_j$ is distributed like a $\chi_2^2$ variable. Therefore, the goodness of fit can be evaluated by applying a Kolmogorov-Smirnov test to the distribution of these ratios.

To generate confidence intervals for the best fit model paramaters, the technique discussed by Cash (1976, 1979) is easily adapted to the case of scaled $\chi^2$–distributions. If $f(\nu;a_1,\ldots,a_p)$ provides a good fit to the sample spectrum, then by using the probability density function in Eq. 5, the likelihood ratio $L$ can be defined

$$L = \frac{max\prod_{j=1}^{n}p_j(y_j,f(\nu_j;a_1^T,\ldots,a_q^T,a_{q+1},\ldots,a_p))}{max\prod_{j=1}^{n}p_j(y_j;f(\nu_j;a_1,\ldots,a_p))} \quad (7)$$

where the maxima are in the parameter space and the $q$ parameters $a_j^T$ are held fixed to their 'true' values. It is possible to demonstrate that for large values of $n$, $-2\log L$ is distributed as $\chi_q^2$, for almost any statistical distribution $p(y;f(\nu;a_1,\ldots,a_p))$. The procedure to generate confidence intervals on model parameters by using the appropriate $-2\log L$ is described by Cash (1976, 1979).

The difference between the standard (least squares) and the maximum likelihood fitting techniques is illustrated in Fig. 2 through an application to the sample power spectra from the X-ray light curve of an EXOSAT observation of the Seyfert Galaxy MCG 6-30-15. The model function consists of the sum of a constant (=2), representing the counting statistics white noise, plus a power law, to model the "red noise" component which is clearly seen. It is apparent that the maximum likelihood fit provides a far more accurate representation of the sample spectrum.

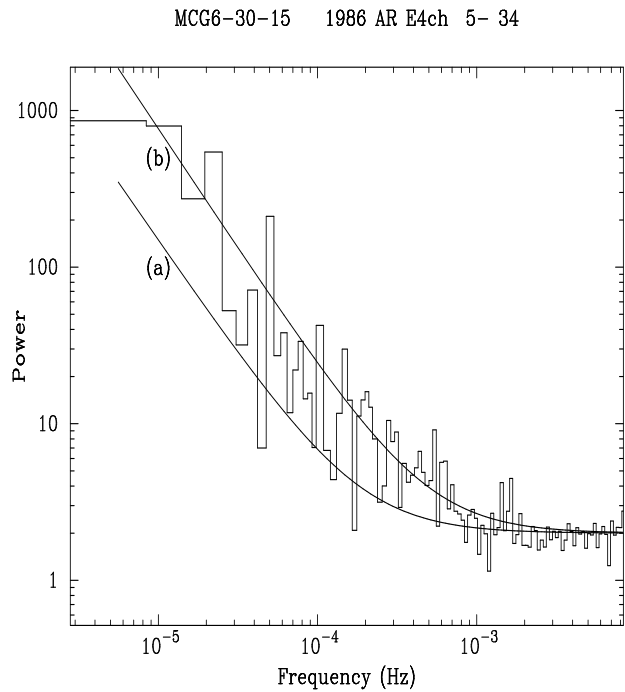

Figure 2: Comparison between the least squares model fitting (a) and maximum likelihood model fitting (b) of the sample power spectrum from a 1-9 keV EXOSAT light curve of the Seyfert galaxy MCG 6-30-15. Note that power estimates from adjacent Fourier frequencies have been averaged so as to make the relative frequency resolution nearly constant.

## 5 Application to the EXOSAT light curves of AGNs

We selected and analysed a sample of long X-ray light curves ($> 20000$ s) of ten bright AGNs obtained from the EXOSAT database (see also Green et al. 1993 and references therein). These are the Seyfert galaxies MCG 6-30-15, NGC 4051, NGC 4151, NGC 4593, NGC 5506, NGC 5548 and MKN 766, the quasar 3C 273, and the BL Lacs MKN 421 and PKS 2155–304. Depending on the signal to noise ratio of the light curves, up to four energy bands were considered: the 0.05–2 keV (LE) light curves obtained from the Channel Multiplier Array in the focal plane of the low energy telescope and used in conjunction with the 3000 Å Lexan filter; the 1–4 keV (A), 4–9 keV (B) and 1–9 keV (C) light curves obtained from the Argon chambers of the Medium Energy proportional counter array, respectively. The sample spectra were calculated over the longest possible interval for each observation by using a direct Fourier transform algorithm. In general the power spectra were dominated by a continuum of $\sim 10^{-3}$ Hz (see e.g. Fig. 2) A model consisting of a power law plus a constant was fitted to each sample spectrum by using the maximum likelihood technique. The power law slopes, $\alpha$, for each AGN and energy band were calculated together with their 90% confidence interval. The application of the new fitting technique shows that the power law slope of the red noise for the AGNs in our sample is systematically steeper than previously estimated, with best fit values ranging between −1.1 and −2.6. The confidence intervals for different observations of the same source are always overlapping, suggesting that the power spectrum slopes are constant in time. A remarkable result is that, contrary to previous claims, only few of the average slopes are marginally consistent with a slope of −1. In particular the average power spectrum slope in each band for all the AGNs in our sample were determined to be: (A) $\alpha \simeq -1.64$, (B) $\alpha \simeq -1.60$, (C) $\alpha \simeq -1.70$ and (LE) $\alpha \simeq -1.69$. The conclusion of this study is that the X–ray variability of the AGNs in our sample over the $\sim 10^{-5} - 10^{-3}$ Hz frequency range is in most cases inconsistent with flicker noise (spectrum power law slope of −1). Most power spectrum slopes are instead consistent with a value of −2, which is characteristic of random walk variability.

## 6 Detecting power spectrum peaks in the presence of "coloured" noise

It is apparent from Sect. 3 that, if $|H(\nu)|^2$ were known, then multiplying Eq. (2) by $|H(\nu)|^{-2}$ the spectrum of the instrumental white noise would be recovered and the search for significant power spectrum peaks could be carried out by using standard techiques. In practice $|H(\nu)|^2$ must be estimated through the sample spectrum. One possibility is to model the power spectrum "coloured" components by using the maximum likelihood technique of Sect. 4 and use the best fit function to estimate $|H(\nu)|^2$. This approach, however, faces difficulties with the subjective choice of the model function and, more crucially, the estimate of the statistical uncertainties of the best fit at any given frequency. Therefore, we preferred to evaluate $|H(\nu)|^2$ through a smoothing algorithm.

As the goal of any power spectrum periodicity search is to detect a signifcant sharp peak, the power in the peak should not affect the estimate of the underlying continuum (otherwise the sensitivity would be reduced). This implies that, for each frequency $\nu_j$, $|H(\nu_j)|^2$ should be estimated through an interpolation of the sample spectrum which excludes $P(\nu_j)$ itself and uses the power estimates over a range of nearby frequencies to the left and the right of $\nu_j$. In the language of the smoothing functions, this corresponds to a class of spectral windows the value of which is zero rectangular window (with a central gap) that extends over a total of $J$ Fourier frequencies, corresponding to a width of $\Delta\nu = J\Delta\nu_F$. In conventional smoothing $(J-1)/2$ Fourier frequencies are used shortwards and longwards of the central frequency $\nu_j$, such that the same smoothing width applies to both sides of the central frequency. The problem with this kind of smoothing is that it does not approximate with acceptable accuracy the steep power–law like red noise components that are often found e.g. in accreting X–ray sources. Fig. 3 shows the results of 1000 simulations of two different types of red power spectra, one consisting of a Lorentzian centered around zero frequency slope of $-1.5$ (spectrum B). In both cases a quasi-periodic oscillation broad peak centered around 100 Hz was included, together with a counting statistics white noise component. A smoothing width of $J = 31$ Fourier frequencies was used. It is apparent that while conventional smoothing (dashed lines in Fig.3) reproduces fairly accurately the characteristics of spectrum A, it fails to reproduce the steep decay from the lowest frequencies of spectrum B. Moreover, edge effects dominate the estimate of the smoothed spectrum for the first $(J-1)/2$ frequencies. A far better result is obtained if the smoothing over $J$ Fourier frequencies is distributed such that its logarithmic frequency width is (approximately) the same on both sides of $\nu_j$, This approach builds on the simple observation that a power law–like spectrum appears like a straight line in a log–log representation (see Israel and Stella 1995 for details).

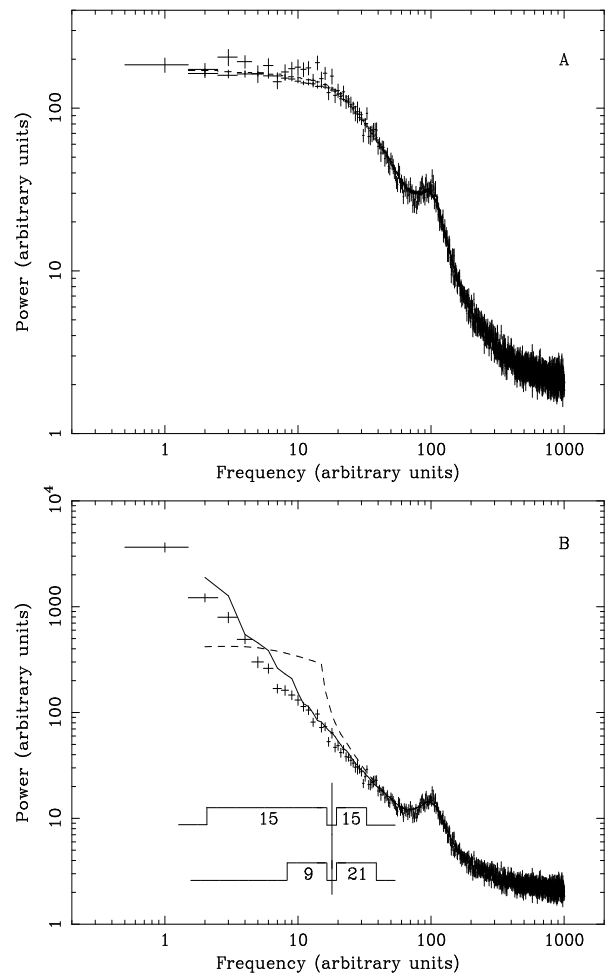

Figure 3: Comparison between the standard rectangular smoothing technique (dashed lines) and the logarithmic interval smoothing technique (solid lines) for the two "coloured" spectral shapes discussed in the text. Lines and points represent the average from 1000 simulations. The smoothing width is $J = 31$. The difference between the two smoothing techniques at a single Fourier frequency is sketched in the bottom panel (the numbers indicate $J_{left}$ and $J_{right}$, i.e. the number of Fourier frequencies used at the left and the right of the nominal frequency $\nu_j$).

The solid lines in Fig. 3 show the estimate of the continuum power spectrum components (and, therefore, of $2|H(\nu)|^2$) obtained by using the above technique; it is apparent that also the low frequency end of spectrum B is reproduced quite well, and that edge effects are virtually absent.

In general, the number of Fourier frequencies in the smoothing ($J = 31$ in the simulations of Fig. 3) is to be adjusted so as to follow closely the sharpest continuous features of the sample spectrum (which tend to favor low values of $L$), while mantaining the noise of the smoothed spectrum as low as possible (which requires high values of $L$). To this aim the divided spectrum $R_j(J) = 2P_j/S_j(J)$ is considered, i.e. the ratio of the sample spectrum and the smoothed spectrum for a range of different values of $J$. If $S_j(J)$ provides a reasonably good estimate of $2|H(\nu)|^2$, then $R_j(J)$, for relatively small values ($R_j(J) \leq 15$ see also Fig. 5), should approximately follow the $\chi_2^2$–distribution of the input white noise. This can be checked with a Kolmogorov–Smirnov (KS) test which compares the integral distributions of $R_j$ and $\chi_2^2$ for a given $J$, and is especially sensitive to differences around the bulk of the distributions (see e.g. Press et al. 1992). The best value of $J$ is assumed to be the one that maximises the KS probability. In practice values of $J$ between $\sim 30$ and the number of Fourier frequencies in the spectrum are used (see below). Fig. 4 shows the results from simulations in which the KS probability is calculated as a function of $J$ for 3 different types of spectra each with 5000 Fourier frequencies. Each point in Fig. 4 represents the average over 100 simulations. The second and third panels refer to spectra B and A of Fig. 3, respectively. In both cases the KS probability shows a broad maximum around values of $J_o \sim 100$. For higher values of $J$ the smoothed spectrum becomes gradually less accurate in reproducing the shape of the sample spectrum, whereas for lower values of $J$ the scatter in the estimates of $S_j(J)$ plays an increasingly important role in distorting the distribution of $R_j(J)$ away from a $\chi_2^2$-distribution. As expected, the KS probability monotonically increases with $J$ in the case of a white noise sample spectrum (see panel W of Fig. 4).

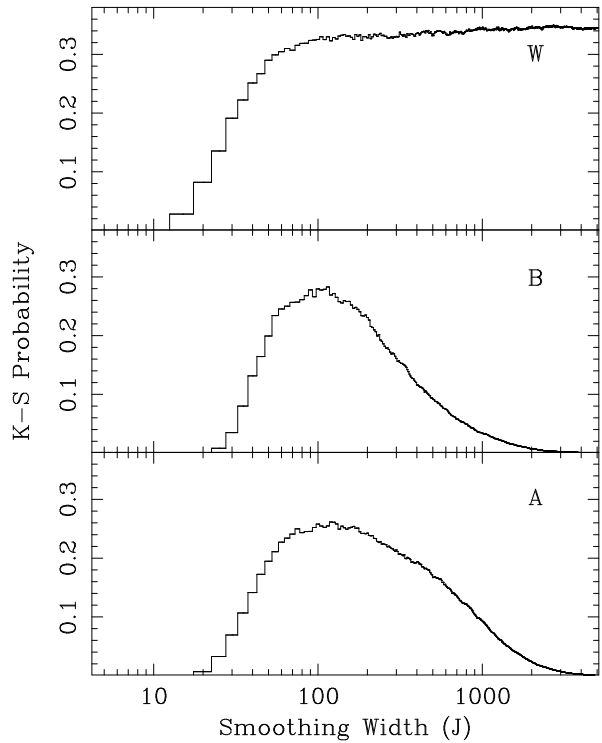

Figure 4: Kolmogorov–Smirnov probability as a function of the smoothing width $J$ for the comparison between a $\chi_2^2$-distribution and $R_j(J)$ in three different cases: a white noise spectrum (W) and the "coloured" spectra of Fig. 3 (A and B). The curves represent the average from 100 simulations.

We adopt the smoothed sample spectrum $S_j(J_o)$ provides the best estimate of $2|H(\nu)|^2$. Therefore, the divided spectrum $R_j(J_o) = 2P_j/S_j(J_o)$ provides our best estimate of the white noise spectrum of the input process. The search for coherent periodicities in the data thus translates in the problem of detecting significant peaks in $R_j(J_o)$. This, in turn, requires a detailed knowledge of the expected distribution of the high values $R_j(J_o)$. For each Fourier frequency $j$, $R_j(J_o)$ is to be regarded as the ratio of the random variables $P_j$ and $S_j(J_o)$. $P_j$ is distributed like a rescaled $\chi_2^2$ with expectation value $S_j(J_o)$ The distribution of $S_j(J_o)$ is in general a suitable linear combination of the $J_o - 1$ $P_j$ random variables used in the smoothing, which, in turn, are distributed like a rescaled $\chi_2^2$. By appealing to the central limit theorem we approximate the distribution of $S_j(J_o)$ with a Gaussian distribution of mean $S_j(J_o)$ and variance $\sigma_j^2(J_o)$ given by the propagation of the variance $2P_j$ of the $P_j$ variables over the smoothing formula. Note that $P_j$ and $S_j(J_o)$ can be regarded, for any given $j$, as statistically independent variables. To check the accuracy and range of applicability of the approximations above, we carried out extensive numerical simulations.

white noise sample spectra each cointaing 5000 Fourier frequencies. The observed distribution of the $R_j(L_o)$ is shown together with the expected distribution derived as outlined above.

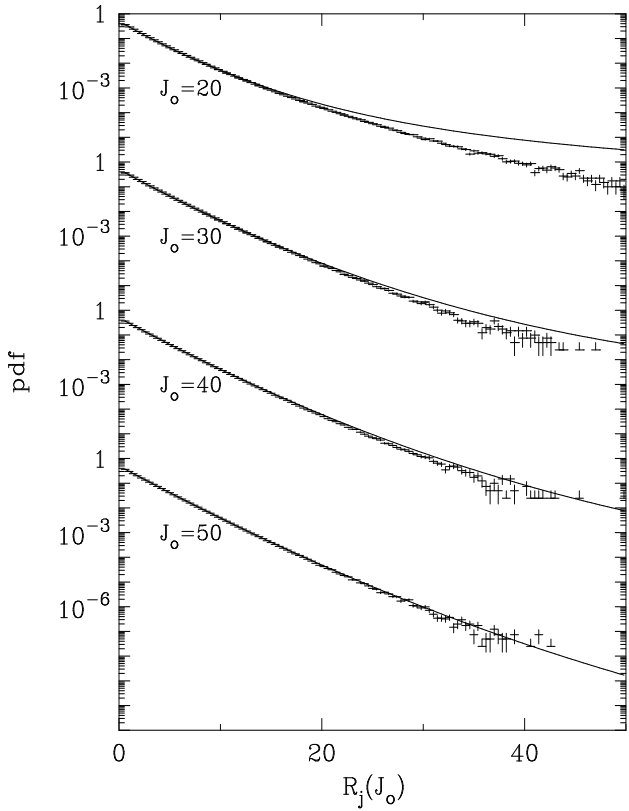

Figure 5: Distribution of $R_j(J_o)$ from $2 \times 10^4$ simulations of a 5000 frequency white noise spectrum, for selected smoothing widths $J_o$. Note that only the power estimates between j=6 and j=4995 were considered. The lines give the expected distribution, calculated as described in the text.

The simulations were repeated for several different choices of the smoothing width $J_o$. It is apparent that while in the cases $J_o = 50$ and 40 the *pdf* in Eq. (3) provides a very good approximation of the observed distribution, for $J_o = 30$ and, especially, $J_o = 20$ the expected *pdf* shows a significant excess for powers larger than $20 - 30$. This effect is due to the fact that the low–value end of the Gaussian approximation for the *pdf* of $S_j(J_o)$ becomes increasingly inaccurate as $J_o$ decreases. Similar results were obtained in the case of white noise spectra obtained from the average of $M$ individual spectra. Also in this case one has to require $MJ_o \geq 40$.

We also tested through extensive simulations the reliability of our approximations for the first few Fourier frequencies of the power spectra (where $J_{left} \ll J_{right}$ because of the logarithmic smoothing intervals) and the frequencies close to the Nyquist frequency (where $J_{left} \gg J_{right}$ because of edge effects). We found that the first and the last 5 − 6 frequencies of the divided spectra cannot be used reliably.

A search for coherent pulsations is then carried out by looking for significant peaks in the divided spectrum, with a probability of chance occurrence below a given detection level. If no significant peaks are found, an upper limit to the semi–amplitude of a sinusoidal modulation is worked out for each Fourier frequency through a generalisation of the method outlined by Leahy et al. (1983).

# 7 A search for orbital X–ray periodicities in Low Mass X–ray Binaries

We have selected and analysed a sample of long EXOSAT lightcurves (from $\sim 2 \times 10^4$ to $\sim 2 \times 10^5$ s) of 25 low mass X-ray binaries (LMXRBs, consisting of a collapsed star accreting from a low mass companion, see van Paradijs 1994) (Table 1). The search was carried out by applying the technique outlined above and by using background subtracted lightcurves from the Medium Energy proportional counters in different energy intervals (1–4, 4–9 and 1–9 keV). X-ray bursts, when present, were removed. Gaps were filled with the average count rate. The power spectra were calculated over the longest possible interval for each observation in order to extend the search to the lowest sampled frequencies.

## 7.1 LMXRBs with Unknown Orbital Period

The shortest orbital period known is that of the globular cluster LMXRB 4U 1820-30 ($\sim$ 685 s, Stella *et al.*, 1987). This is close to the minimum orbital period predicted for a LMXRB with a He mass–donor. Therefore, for the LMXRBs in our sample with unknown orbital period, we carried out a search for sinusoidal modulations with periods ranging from 500 s to $\sim$ 11 hr for GX349+2 and GX9−1, to $\sim$ 2.8 hr for GX13+1, X1705−44 and EXO1747−21 and to $\sim$5.7 hr for the remaining 15 sources. The self–consistent determination of the smoothing width provided values of $J_o$ between $\sim$50 and $\sim$3000 Fourier frequencies. No significant peaks were found with a 95% confidence level in any of the power spectra. The corresponding upper limits to the semi–amplitude of a sinusoidal modulation were worked out for each source in the three different energy intervals. The results for the 1–10 keV band are shown in Figs. 6 and 7.

The curves giving the upper limits reflect the shape of the noise components of the power spectra. In the case of those sources which are faint (5–15 cts/s) and/or do not show a strong red–noise variability, such as the globular cluster sources Terzan1, Terzan2, NGC6712 and NGC1851 and the two sources 2S1715−32 and 2S1543−62, the sensitivity of the search, for frequencies higher than $2 \times 10^{-4}$ Hz, is affected mainly by counting statistics noise and upper limits of ≤0.7% are obtained for any sinusoidal modulation. For lower frequencies an increasing noise component is present in the power spectra and the upper limits reach values around 1–2%.

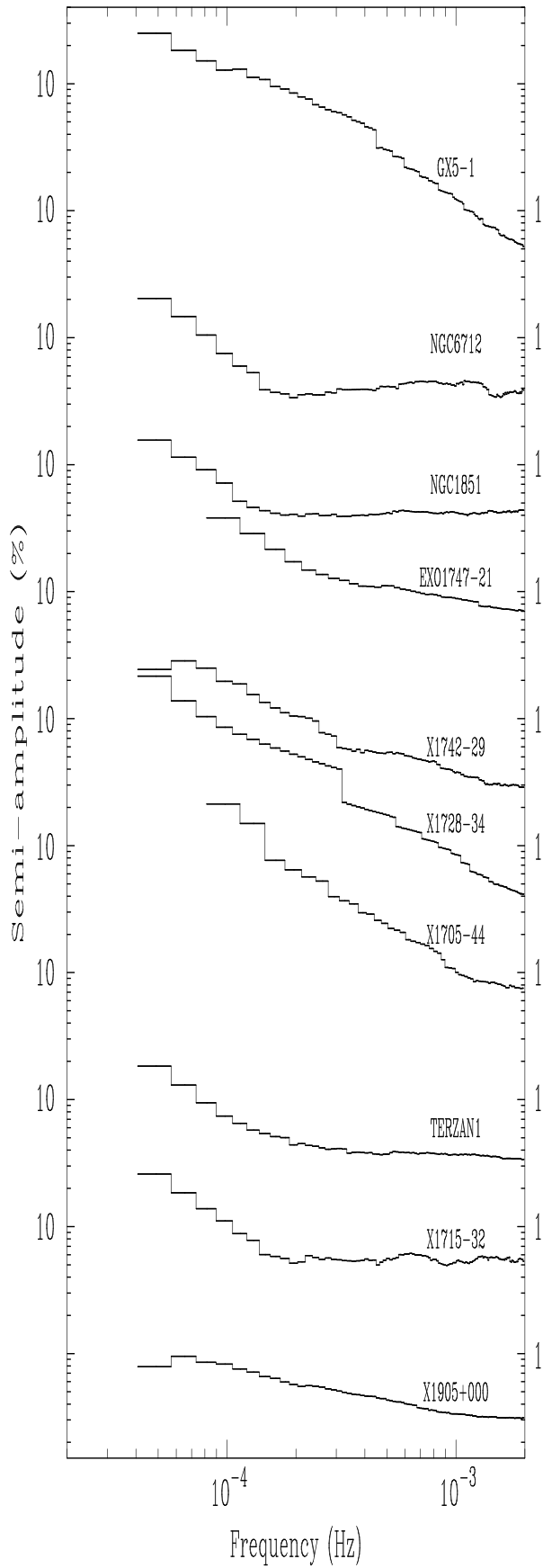

Figure 6: 95% confidence upper limits to the semi–amplitude of a sinusoidal modulation for ten LMXRBs with unknown orbital period.

Table 1: List of LMXRBs

| Name | Time yy.ddd | Expos hr | Count rate cts/s |
|---|---|---|---|
| LMCX-2 | 85.344 | 16.5 | 27.8±0.1 |
| 4U1626-67 | 83.242 | 6.5 | 43.6±0.2 |
| MXB1636-53 | 84.127 | 19.5 | 121.3±0.4 |
|  | 85.219 | 76.5 | 128.3±0.4 |
|  | 85.248 | 24.0 | 181.8±0.4 |
| GX9+9 | 83.269 | 18.2 | 472.1±1.3 |
| MXB1735-44 | 85.227 | 88.7 | 277.8±0.8 |
| NGC1851 | 86.004 | 12.6 | 5.0±0.1 |
| 4U0614+09 | 86.043 | 15.2 | 179.2±0.4 |
| 2S1543-624 | 84.238 | 13.2 | 73.7±0.2 |
| 4U1556-60 | 85.107 | 18.9 | 28.4±0.1 |
| GX340+0 | 85.088 | 9.3 | 677.3±1.5 |
| GX349+2 | 85.243 | 46.9 | 1190.2±2.6 |
| 4U1705-44 | 85.276 | 15.7 | 177.5±0.4 |
| 2S1715-321 | 84.221 | 8.6 | 9.5±0.1 |
| Terzan2 | 85.073 | 12.0 | 11.8±0.1 |
| MXB1728-34 | 84.257 | 17.3 | 38.1±0.2 |
| Terzan1 | 85.253 | 12.2 | 13.9±0.1 |
| 2S1742-294 | 85.103 | 9.3 | 79.9±0.2 |
| EXO1747-21 | 85.101 | 6.7 | 85.2±0.3 |
| GX9+1 | 85.263 | 25.0 | 915.2±2.0 |
| GX5-1 | 85.119 | 10.3 | 1644.0±3.6 |
| GX13+1 | 85.122 | 8.1 | 467.0±1.0 |
| GX17+2 | 83.215 | 7.5 | 900.1±2.8 |
| 2A1822-000 | 85.249 | 18.1 | 50.9±0.1 |
| NGC6712 | 85.261 | 12.5 | 7.5±0.1 |
| 2S1905+000 | 85.251 | 17.2 | 18.4±0.1 |

For all the other sources a pronounced power law–like, red noise component plays an increasingly important role in reducing the sensitivity of the search. The corresponding upper limits range from ∼3% to 50% for periods between 2.8 and 11.4 hr.

## 7.2 LMXRBs with Optical Orbital-Modulation

Using the EXOSAT database we selected (see Table 1) five LMXRBs, the orbital period of which has so far been detected only at optical wavelengths (except for GX9+9). The presence of additional red noise due to the aperiodic variability of these sources has hampered the detection of X-ray modulations at the optical period. The search for a sinusoidal X-ray modulation was carried out on the basis of the new technique over the narrow interval of Fourier frequencies allowed by the period measurement in the optical band.

The power spectra from the 1-4, 4-10 and 1-10 keV lightcurves were calculated for LMC X−2 ($P_{orb}$=6.4 hr), 4U 1626−67 ($P_{orb}$=0.7 hr), MXB 1636−539 ($P_{orb}$=3.8 hr), MXB1735−44 ($P_{orb}$=4.6 hr) and GX9+9 ($P_{orb}$=4.2). Best smoothing widths between

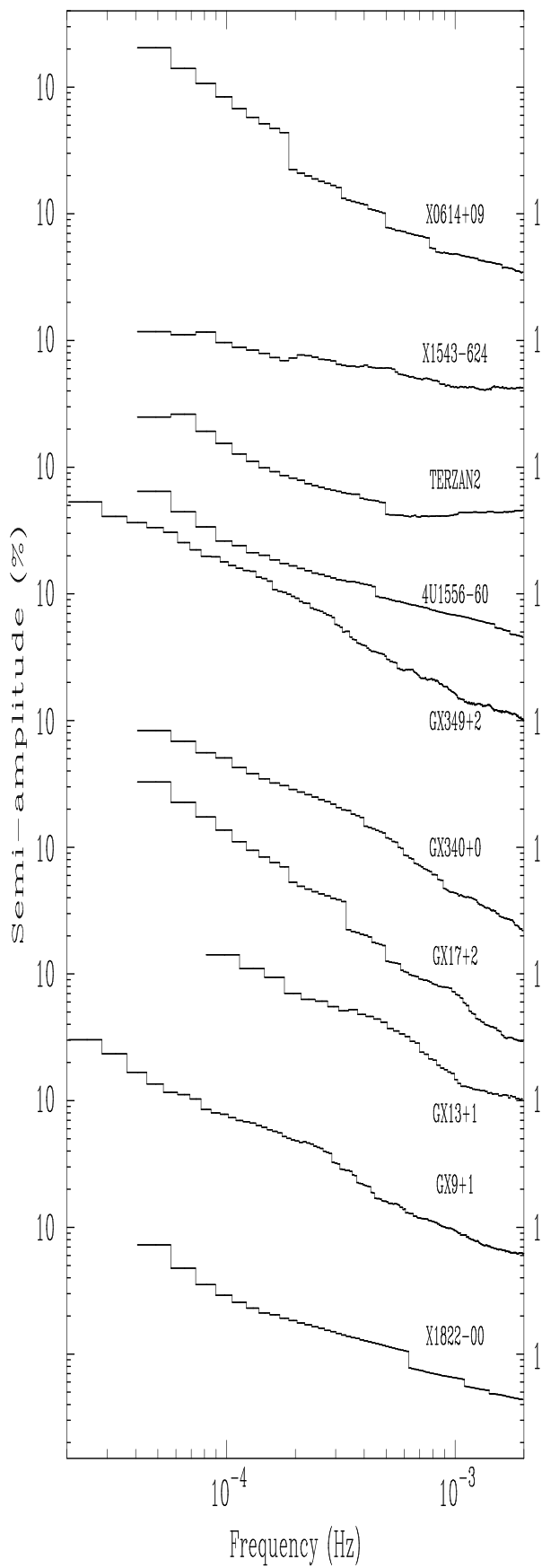

Figure 7: 95% confidence upper limits to the semi–amplitude of a sinusoidal modulation for the other ten LMXRBs with unknown orbital period.

Table 2: Results from the search for sinusoidal modulations in the 1–10 keV lightcurves of LMXRBs with known orbital period. The best probability values obtained in different energy intervals are reported.

| Source | $P_{orb}$ (hr) | $A_{UL}$ (%) | A (%) | Prob. (%) | E* |
|---|---|---|---|---|---|
| LMC X-2 | 6.4 | — | $1.2\pm^{1.2}_{0.5}$ | 90.4 | C |
|  |  |  | $1.4\pm^{0.7}_{0.8}$ | 92.8 | B |
| X1626−67 | 0.7 | 2.2 | — | — | C |
| X1636−53* | 3.8 | — | $1.5\pm^{0.7}_{1.1}$ | 95.8 | C |
|  |  |  | $1.4\pm^{0.7}_{0.5}$ | 98.5 | A |
| X1636−53† |  |  | $4.3\pm^{1.4}_{2.1}$ | 98.6 | C |
|  |  |  | $1.9\pm^{0.8}_{1.6}$ | 99.8 | A |
| X1636−53‡ |  | 4.3 | — | — | C |
| GX9+9 | 4.2 | — | $0.8\pm^{0.5}_{0.2}$ | 93.3 | C |
|  |  |  | $3.0\pm^{0.3}_{0.3}$ | 99.7 | B |
| X1735−44 | 4.6 | 2.0 | — | — | C |

* Energy: A(1–4 keV), B(4–10 keV) and C(1–10 keV).
* 1984 day 127 observation.
† 1985 day 248 observation.
‡ 1985 days 219–221 observation.

∼40 and ∼1200 independent Fourier frequencies were determined.

In the case of LMC X−2, during the 17 hour observation of 1985 December 10, a low amplitude modulation was found at 93% confidence level (see Table 2) in a search at the frequency corresponding to optical modulation.

We searched for a significant power spectrum peak at ∼4× $10^{-4}$ Hz in a 7.6 hour EXOSAT observation of 4U 1626−67. No significant X−ray modulation was found with a 95% confidence upper limit of ∼2%.

The search was also carried out for the three longest EXOSAT observations of MXB1636−539. A significant peak was detected during the 17 hr 1984 May 8 and the 23 hr 1985 September 6 observations with a (maximum) significance of 98.5% and 99.8% respectively. The corresponding semi–amplitude of the X−ray modulation was of about 2-4% (see Table 2 and Fig. 8). No significant peaks were detected instead in the power spectra from the 59 hour 1985 August 7–9 observation: due to the strong aperiodic variability of the source a ∼4% upper limit was obtained for the semi–amplitude of a modulation at the orbital period. This value is larger than the semi–amplitude measured during the other two observations.

MXB 1735−44 was observed for 3 days in 1985 August 16–18 by EXOSAT. No significant peak at the optical orbital period of ∼4.6 hr was found in the power spectra. The upper limits determined with the new technique are about ∼2%.

A 4.2 hr orbital modulation from GX 9+9 was discovered in the X−ray band using the HEAO−1 data (Hertz and Wood, 1988). In order to confirm the existence of this signal we analysed the lightcurves from the 9.2 hr 1983 September 27 observation. We found a significant (up to 99.7%) peak in the power spectra

This work was partially supported through ASI grants.

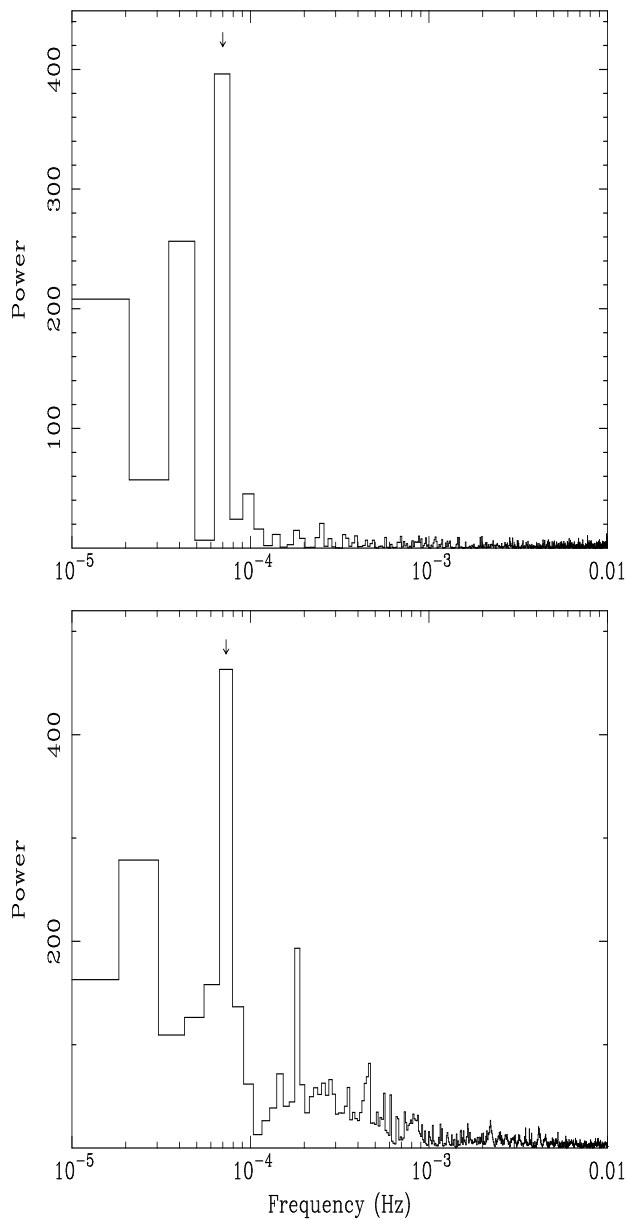

Figure 8: Power spectra from the 1984 day 127 (upper panel) and 1985 day 248 (lower panel) observations of MXB 1636−539. The arrows mark the frequency corresponding to the orbital period of 3.8 hr.